\documentclass[
  ,draft            % use draft while you are working on the paper
  ]
  {aipproc}

\layoutstyle{8x11single} %% kept this as it was in original (for
\begin{document}

\title{Type Ia Supernovae and Accretion Induced Collapse}

\classification{90}
\keywords      {Supernovae -- White Dwarfs -- Neutron Stars}

\author{A. J. Ruiter}{
  address={Max Planck Institute for Astrophysics,
Karl-Schwarzschild-Str. 1, 85741 Garching, Germany}
}

\author{K. Belczynski}{
  address={Los Alamos National Laboratory, CCS-2/ISR-1 Group, P.O. Box
1663, Los Alamos, NM 87545 USA; Oppenheimer Fellow}
,altaddress={Astronomical Observatory, University of Warsaw, Al. Ujazdowskie
4, 00-478 Warsaw, Poland} % additional visiting address
}

\author{S. A. Sim}{
  address={Max Planck Institute for Astrophysics,
Karl-Schwarzschild-Str. 1, 85741 Garching, Germany}
}

\author{W. Hillebrandt}{
  address={Max Planck Institute for Astrophysics,
Karl-Schwarzschild-Str. 1, 85741 Garching, Germany}
}

\author{M. Fink}{
  address={Max Planck Institute for Astrophysics,
Karl-Schwarzschild-Str. 1, 85741 Garching, Germany}
}

\author{M. Kromer}{
  address={Max Planck Institute for Astrophysics,
Karl-Schwarzschild-Str. 1, 85741 Garching, Germany}
}

\begin{abstract}

Using the population synthesis binary evolution code {\tt StarTrack}, 
we present theoretical rates and delay times of Type Ia supernovae
arising from various formation channels.  These channels include
binaries in which the exploding white dwarf reaches the
Chandrasekhar mass limit (DDS, SDS, and helium-rich donor scenario) as
well as the sub-Chandrasekhar mass scenario, in which a white dwarf 
accretes from a helium-rich companion and explodes as a SN Ia
before reaching the Chandrasekhar mass limit. 
We find that using a common envelope parameterization employing
energy balance with $\alpha_{\rm CE}=1$ and $\lambda=1$, 
the supernova rates per unit mass (born in stars) 
of sub-Chandrasekhar mass SNe Ia exceed those of all other 
progenitor channels at epochs $t=0.7 - 4$ Gyr for a burst of star
formation at $t=0$.  Additionally, the delay time distribution of 
the sub-Chandrasekhar model can be divided in to two
distinct evolutionary channels: the `prompt' helium-star channel 
with delay times $< 500$ Myr, and the `delayed' double white dwarf
channel with delay times $> 800$ Myr spanning up to a Hubble
time.  These findings are in agreement with recent
observationally-derived delay time distributions which predict that a
large number of SNe Ia have delay times $< 1$ Gyr, with a
significant fraction having delay times $< 500$ Myr.  
We find that the DDS channel is also able to account for the
observed rates of SNe Ia.  However, detailed simulations of white dwarf
mergers have shown that most of these mergers will not lead to SNe Ia 
but rather to the formation of a neutron star via accretion-induced
collapse.  If this is true, our standard population synthesis 
model predicts that the only progenitor channel which can 
account for the rates of SNe Ia is the sub-Chandrasekhar mass
scenario, and none of the other progenitors considered can fully account
for the observed rates.

\end{abstract}

\maketitle

%%%%%%%%%%%%%%%%%%%%%%%%%%%%%%%%%%%%%%%%%%%%
%% MAINMATTER
%%%%%%%%%%%%%%%%%%%%%%%%%%%%%%%%%%%%%%%%%%%%

\section{Introduction}

The progenitors of Type Ia supernovae (SNe Ia)  
remain unknown, though the idea that SN Ia progenitors belong to at least two distinct
populations \citep[see e.g.,][]{SB05,MDP06,MB10,Bra10} has recently been gaining
ground.  A picture is emerging 
which supports populations of both quickly-evolving (prompt) 
progenitors with short delay times less than $t \sim 500$ 
Myr, as well as more slowly-evolving progenitors with 
(sometimes rather) long delay times spanning up to a Hubble time.

The most favoured SN Ia progenitor scenarios involve
the double degenerate scenario (DDS; \citet{Web84}), and the single 
degenerate scenario (SDS; \citet{WI73}).  In the DDS, the merger 
of two carbon-oxygen (CO) white dwarfs (WDs) with a total mass exceeding the Chandrasekhar mass limit, 
$M_{\rm Ch} \sim 1.4$ M$_{\odot}$, can lead to 
explosive carbon-burning which causes a SN Ia explosion.  
In the SDS, a CO WD accretes from a non-degenerate 
stellar companion via stable Roche-lobe overflow (RLOF),
enabling the WD to accumulate mass toward $M_{\rm Ch}$ 
until carbon is ignited explosively in the centre leading to a SN Ia.  
The companion filling its Roche-lobe can be hydrogen-rich 
or helium-rich; in this work we delineate between SDS
(hydrogen-rich donors) and helium-rich donor cases, as in
the latter case the donor may be a non-degenerate helium-burning star, 
or a degenerate helium-rich WD.

For some time, population synthesis calculations
\citep[e.g.,][]{Yun94} 
have predicted that the number of merging CO WDs (DDS) with a
total mass exceeding 
$M_{\rm Ch}$ is sufficient to match, and thus possibly account for, the rate of SNe Ia 
\citep[$0.4 \pm 0.2$ per century for the Galaxy,][]{CET99}.  
At the same time, despite the fact that potential SDS progenitors have
been observed \citep{HK01}, the theoretically-predicted SN Ia rate
from the SDS channel is usually unable to explain the observed rates 
of SNe.  In the majority of population studies, the relative frequency of SDS events 
is often found to be well below those of the DDS \citep[][see also \citet{Men10}]{YL98,RBF09} 
being about an order of magnitude too low compared to observations 
of SNe Ia.

Detailed WD merger simulations have shown that some WD mergers can
successfully lead to a SN Ia explosion \citep{Pie03,Pak10}.  However, 
the main argument against the DDS is that 
the merging process leads to physical conditions 
in which a thermonuclear explosion is unlikely \citep{YPR07}.  
It is more probable that a merger between two CO WDs with a total mass 
$> M_{\rm Ch}$ will collapse and form a neutron star; an accretion 
induced collapse \citep[AIC,][]{Miy80}.  Such events could be
detectable in modern transient surveys, though would 
have an observational signature unlike those of SNe
Ia \citep{Dar10}. 

Recently, \citet{RBF09} carried 
out a population synthesis study showing rates and delay 
times for 
SNe Ia involving WDs whose mass at the time of SN Ia 
had reached the Chandrasekhar mass limit: DDS, SDS and 
helium-rich donor scenarios (`AM CVn' channel in that work).  
We extend our investigation of progenitors and 
focus on the sub-Chandrasekhar mass model (detailed
results are presented in Ruiter et al. 2010, in preparation).  
Since these calculations are based on the work which was performed for
\citet{RBF09}, the reader is referred to that paper for a more
detailed description of the DDS, SDS and helium-rich (Chandrasekhar
mass WD) donor scenario.

\subsection{Sub-Chandrasekhar SNe Ia}

Sub-Chandrasekhar SNe Ia -- SNe Ia which take place in WDs whose 
mass is below $M_{\rm Ch}$ -- likely consist of a (probably CO) 
WD accreting via stable RLOF from a helium-rich companion \citep{IT91}.
These systems have thus far been regarded 
as an unlikely model for SNe Ia owing to the fact that most synthetic 
light curves and spectra of these objects from previous studies did
not match those of observations \citep[e.g.,][]{HK96,Nug97}, likely owing to
the thickness of the helium shell ($\sim 0.2 M_{\odot}$).  
More recently, \citet{SB09} have shown that conditions 
suitable for a detonation in the WD might be achieved for even lower
helium shell masses than were assumed in previous studies.
Double detonations in sub-Chandrasekhar mass WDs with 
low-mass helium shells have shown to be a robust explosion mechanism
that can produce SNe Ia of normal brightness, provided that a detonation in
the helium shell is successfully triggered \citep{Fin10}.  
Recent results involving hydrostatic sub-Chandrasekhar mass 
exploding WDs with subsequent nucleosynthesis and radiative 
transfer calculations \citep{Sim10,Kro10} indicate that the sub-Chandrasekhar mass 
scenario should be considered as a likely SN Ia progenitor 
candidate.  Additionally, population synthesis studies \citep{RBF09} predict that 
there are a sufficient number of these binaries to explain the observed rate of 
SNe Ia.

\section{Model Description}

We use the {\tt StarTrack} population synthesis binary evolution code \citep{Bel08} to 
evolve our stellar population.    
%Many of the updates concerning 
%accretion on to WD systems can be found in \citet{BBR05,Bel08}, though since then we have 
%incorporated an updated prescription for accretion of hydrogen on WDs \citep{Nom07}.  
The initial distributions for binary orbital parameters (orbital periods, mass ratios, etc.) 
are the same as described in \citet[][section 2]{RBF09}.  
In \citet{RBF09}, it was assumed that the ejection of the envelope of the mass-losing star
during a CE phase came at the expense of removing the orbital energy of the 
binary, as dictated by the `energy-balance' equation (or `$\alpha$-formalism') 
\citep{Web84}
\begin{equation}
\frac{G \, M_{\rm don,i} \, M_{\rm ej}}{\lambda \, R_{\rm don,i}} = \alpha_{\rm CE}\left( \frac{G \, M_{\rm don,f} \, M_{\rm com}}{2 \, a_{\rm f}} - \frac{G \, M_{\rm don,i} \, M_{\rm com}}{2 \, a_{\rm i}} \right)
\end{equation}
where $G$ is the gravitational constant, $M_{\rm don,i}$ is the initial mass of the (giant) 
donor star just prior to the CE, $M_{\rm ej}$ is the ejected mass
(assumed to be the mass of the giant's envelope), 
$M_{\rm com}$ is the mass of the companion (assumed to be unchanged during the CE), $M_{\rm don,f}$ 
is the final mass of the donor once the envelope has been ejected, $R_{\rm don,i}$ is the initial 
radius of the donor star when it fills its Roche-lobe, $a_{\rm i}$ is the initial orbital  
separation, $a_{\rm f}$ is the final orbital separation (if $a_{\rm f}$ is too small to 
accommodate the Roche-lobes of the stars, the CE results in a merger), 
$\alpha_{\rm CE}$ is the efficiency with which the binary orbital energy 
can unbind the CE, and $\lambda$ is a parameterization of the structure of the 
donor star; both $\alpha_{\rm CE}$ and $\lambda$ are
fairly uncertain.  Here we employ $\alpha_{\rm CE}=1$ and $\lambda=1$. 
In an upcoming paper we investigate the impact 
of the CE parameterization on SN Ia delay times and rates 
(Ruiter et al. 2010, in preparation).

\subsection{Sub-Chandrasekhar Mass Model}

It has been shown previously that a WD accumulating 
helium-rich material may be capable of exploding in a Type Ia supernova if the 
correct conditions are satisfied; even if the WD is below the Chandrasekhar 
mass limit \citep{Taa80,IT91,WW94,IT04}.  
We adopt the prescription of \citet{IT04}, applied to 
accretion from helium-rich companions only, to determine when a particular 
binary undergoes a sub-Chandrasekhar mass SN Ia \citep[][see section 5.7.2
  for equations]{Bel08}. 
In short, we consider three different accretion rate regimes for 
accumulation of helium-rich material on CO WDs, adopting the
input physics of \citet{KH99,KH04}.  
At high (e.g., initial RLOF) accretion rates, helium 
burning is stable and thus mass accumulation on the WD is fully efficient 
($\eta_{\rm acu} = 1$).  At somewhat lower accretion rates, helium burning is unstable and 
the binary enters a helium-flash cycle, thus accumulation is possible but is not fully 
efficient ($\eta_{\rm acu} < 1$).  In both of these aforementioned accretion regimes, 
the CO WD is allowed to accrete (and burn) helium, and its total mass
may reach the Chandrasekhar mass limit within a Hubble time and
explode as a Type Ia supernova through the helium-rich donor channel.  
However, for low accretion rates 
%\citep[e.g., log$\dot{M}_{\rm He} $($M_{\odot}$yr$^{-1}$)$ < -7.6$ for
%  WD masses $< 0.7 M_{\odot}$,][]{KH04}, 
compressional heating at the base of the accreted helium layer 
plays no significant role, and a layer of unburned helium can be accumulated on the WD surface.  Following 
\citet{IT04}, we assume that if such a CO WD accumulating helium enters this `low' accretion rate 
regime and accumulates $0.1 M_{\odot}$ of helium on its surface, a
detonation is initiated at the base of the helium shell layer.  Consequently, a detonation in 
the core of the CO WD is presumed to follow, and we assume that a
sub-Chandrasekhar Type Ia supernova takes place. 
In our model, only accreting WDs with a {\em total mass} $> 0.9 M_{\odot}$ are considered to 
lead to potential sub-Chandrasekhar SNe Ia, since lower mass cores,
while they may detonate, are unlikely to produce enough radioactive nickel and
hence will not be visible as SNe Ia.  Thus in all future
discussions we refer to sub-Chandrasekhar systems whose total WD mass
(CO core $+$ helium shell) 
is at least $0.9 M_{\odot} $ at the time of SN Ia.

\section{Results}

For an instantaneous burst of star formation at $t=0$,  
we have calculated the DTD and rates 
of SNe Ia arising from different formation channels: 
the DDS, SDS, helium-rich donor, and the sub-Chandrasekhar scenario. 
We find that only two SN Ia formation scenarios are capable of  
matching the observed SNe Ia rates: 
the DDS and the sub-Chandrasekhar channels.
The adopted sub-Chandrasekhar scenario ($M_{\rm WD} > 0.9$) 
is dominant at nearly all epochs $\leq 5$ Gyr. 
As found in \citet{RBF09}, the DDS DTD follows a continuous 
power-law with the largest number of events occurring starting at $\sim 50$
Myr. We present for the first
time the DTD of the sub-Chandrasekhar mass
channel, which is clearly divisible into prompt and a delayed
components.  

\subsection{Delay Times}

\begin{figure}[tpb]
    \includegraphics[height=.5\textheight]{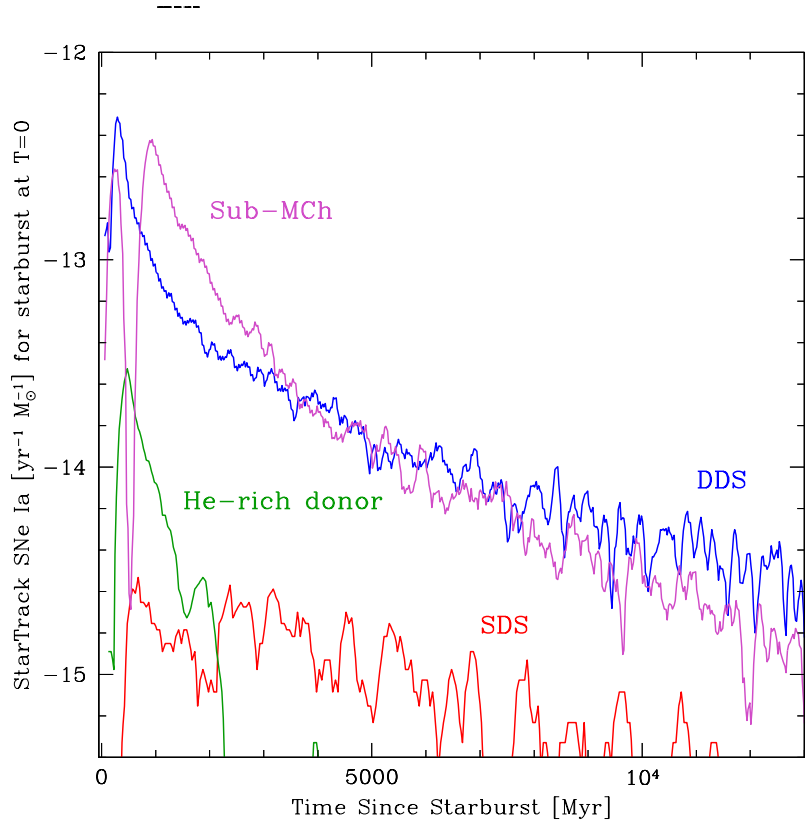}
\caption{
Delay time distribution. The number of SNe Ia 
per year per unit stellar mass born in stars (at 
starburst t=0, 50\% binarity) is shown for the four formation channels
considered: DDS, SDS, helium-rich donor,
Chandrasekhar mass WD, and sub-Chandrasekhar.  
Despite the different scaling on the $y$-axis from \citet{RBF09}, 
the shapes of the DTDs from the three Chandrasekhar (or above) mass models are the same as in
that work (Model 1).  The newly-calculated sub-Chandrasekhar SN Ia DTD
clearly shows two distinct populations: the helium star channel (spike
at delay times less than $\sim 500$ Myr) and the helium WD channel
(from $\sim 800$ Myr to a Hubble time).  
}
\label{fig:dta1a125g1}
\end{figure}

In Figure~\ref{fig:dta1a125g1} we show the DTD of
the four progenitor channels investigated in this work.  
We show the delay times in units of SNe Ia per year 
per unit stellar mass born in stars 
(SNuM $\times 10^{-12}$; SNuM = SNe Ia per century per $10^{10} M_{\odot}$). 
For our DTD normalization, we have assumed a binary fraction of 50\% 
across the entire initial stellar mass function.  
We note that the bumps in the smoothed plot are
due to Monte Carlo noise.

As was found in \citet{RBF09} Model 1, 
the DDS DTD follows a power-law shape with $\sim$ $t^{-1}$,
while the SDS distribution is somewhat flat with no events with delay times
less than $\sim 450$ Myr.  The reason why the SDS does not harbour prompt
events is directly linked to the donor star's ZAMS mass; all SDS
progenitor donors are found to have ZAMS masses below $2.8 M_{\odot}$.  When the 
secondary ZAMS mass is $>2.8 M_{\odot}$, the binary will enter a 
CE phase when the secondary fills its Roche-lobe, rather than a stable
RLOF phase.  In such a case, the binary will not become an
SDS SN Ia, though may under the right circumstances evolve to SN Ia 
from the helium-rich donor channel.  The helium-rich donor DTD 
consists mostly of systems with relatively short ($\sim$100 Myr $-2$
Gyr) delay times, with very few events at long delay times. We refer
the reader to \citet{RBF09} for a description of these DTDs;
The overall DTD shapes remain the same as in that work. 

The sub-Chandrasekhar mass systems can easily by eye be grouped into two 
classes: those which occur with delay times less than $\sim 500 $ 
Myr, and those with delay times above $\sim 800$ Myr.
Not surprisingly, these two classes of SNe Ia 
stem from two very different evolutionary scenarios.  Those with short 
delay times 
consist of progenitors which involve a helium-burning star donor, 
where as the rest consist of helium or hybrid (CO core, helium mantle)  
WD donors.  

The prompt component (delay times $< 500$ Myr) accounts for $\sim 13$ \% of all
sub-Chandrasekhar SNe Ia that explode within 13 Gyr of star
formation. Nearly all of these systems ($\sim 10$\% of the total
sub-Chandrasekhar SNe) have helium star donors, with a
small fraction of the prompt component originating from the hybrid 
WD channel. The delay time is governed by the main sequence (MS) lifetime of the
donor star.  The companions with ZAMS masses $\geq 3 M_{\odot}$
evolve off of the MS within $\leq 400$ Myr.  After the first CE,
which leaves behind a CO primary WD and a MS secondary star, the
(slightly evolved) secondary will fill its Roche-lobe 
and mass transfer is once again unstable leading to a second CE
phase.  The CE brings the CO WD and newly-formed naked helium 
star on a close orbit ($\sim 35-40$ min. orbit).  Within $\sim$ a few Myr the helium
star fills its Roche-lobe.  However, initial mass transfer rates for
the helium star channel fall within the `low accretion rate regime': 
typically such systems have initial mass transfer rates 
$\sim 2 \times 10^{-8} M_{\odot}$yr$^{-1}$.

The delayed component (delay times $> 500$ Myr) comprise the other
$\sim 87 $ \% of the sub-Chandrasekhar mass progenitors; binaries with
helium WD donors make up $\sim 78$ \% (the rest being hybrid WD donors).  These binaries
also evolve through two CE phases, as is expected for the evolution of
AM CVn binaries.  Similar to the DDS, the timescale governing the DTD
for the helium WD channel is primarily set by the gravitational radiation
timescale.  However unlike the DDS, these WDs 
do not merge upon contact, but enter
a stable phase of RLOF.  

\subsection{Rates}

In Table 1, we show the SN Ia rates for a  
spike of star formation at $t=0$, 
and a binary fraction of 50\%.    
As was determined in \citet{RBF09}, Model 1, DDS rates are able to 
(just) account for the observed Galactic rate of SNe Ia, where as both the 
SDS and helium-rich donor channels fall short by over an order of magnitude.  
We find that the rate of our adopted sub-Chandrasekhar SN Ia 
model exceeds all other progenitor channels between $\sim 0.7-4$ Gyr,
and these systems are enough to account for the Galactic SN Ia 
rate, with a calculated Galactic rate of $\sim 2.6 \times 10^{-3} $ SN
Ia yr$^{-1}$.  For comparison, the DDS Galactic rate is 
$\sim 2 \times 10^{-3} $ SN Ia yr$^{-1}$ 
\citep[see][section 4 for discussion of Galactic rate
  calculation]{RBF09}.  
Both of these values are within the Galactic rate estimate 
from \citet[][]{CET99} of $4 \pm 2 \times 10^{-3} $ SN Ia yr$^{-1}$ 
\citep[see also][for a modern SN rate calculation in the local Universe]{Li10}. 

\begin{table}[tbp]
\begin{tabular}{lrrrr}
\hline
\tablehead{1}{l}{b}{Time}
& \tablehead{1}{c}{b}{Rate}\\
%  & \tablehead{1}{r}{b}{Total}   \\
\hline
DDS        & \\
 0.1 Gyr & $2.0 \times 10^{-1}$ \\
 0.5 Gyr & $1.6 \times 10^{-1}$ \\
  1 Gyr  & $8.0 \times 10^{-2}$ \\
  3 Gyr  & $2.5 \times 10^{-2}$  \\
  5 Gyr  & $1.2 \times 10^{-2}$\\
 10 Gyr  & $\sim 5 \times 10^{-3}$ \\
 & \\
SDS                    &  \\
 0.1 Gyr & $ 0 $ \\
 0.5 Gyr & $\sim 10^{-3}$ \\
  1 Gyr  & $1.5 \times 10^{-3}$ \\
  3 Gyr  & $2.0 \times 10^{-3}$ \\
  5 Gyr  & $\sim 1 \times 10^{-3}$ \\
 10 Gyr  & $\leq 10^{-3}$\\
 & \\
Helium donor ($M_{\rm Ch}$)             & \\
 0.1 Gyr & $\sim 3 \times 10^{-3}$ \\
 0.5 Gyr & $ 2.2 \times 10^{-2}$ \\
  1 Gyr  & $ 8.0 \times 10^{-3}$ \\
  3 Gyr  & $ < 10^{-3}$ \\
  5 Gyr  & $ \leq 10^{-4}$\\
 10 Gyr  & $ \sim 0 $ \\
 & \\
sub-Chandrasekhar             & \\
 0.1 Gyr & $\sim 1 \times 10^{-1}$ \\
 0.5 Gyr & $ \sim 10^{-3}$ \\
  1 Gyr  & $ 3.3 \times 10^{-1}$ \\
  3 Gyr  & $ 4.0 \times 10^{-2}$ \\
  5 Gyr  & $ 1.4 \times 10^{-2}$\\
 10 Gyr  & $ \sim 4 \times 10^{-3} $ \\
\hline
\end{tabular}
\caption{Rates of SNe Ia (SNuM) following a
      starburst at t=0.}
\label{tab:a}
\end{table}

\section{Discussion}

Hydrodynamic simulations of sub-Chandrasekhar 
mass SNe Ia exhibit observational features which are
characteristically similar to those of SNe Ia 
\citep{Sim10,Kro10}.  Motivated by these 
findings, as well as population synthesis rate 
estimates, we have investigated sub-Chandrasekhar 
mass SN Ia formation channels and have calculated 
and presented the DTDs and rates of their
progenitors.  

Within the framework of our adopted model, 
we find that only the DDS and sub-Chandrasekhar mass channels
can potentially explain the observed rates of SNe Ia.  
However, while the predicted rates of the 
DDS are not in conflict with observations, these systems are
theoretically expected to produce events which lead to AIC. 
If this is the case, the Galactic AIC rate from this AIC 
evolutionary (WD merger) channel alone would be as high as 
$\sim 10^{-3}$ per year.  An
absence of DDS progenitors from the SN Ia DTD would leave the
sub-Chandrasekhar mass model as the only evolutionary scenario 
able to produce a high enough rate of SNe Ia, as well as a bimodal DTD.  
However, if one can say for certain that AICs
which are formed from the merger of CO WDs produce very neutron-rich
ejecta\footnote{\citet{YL98} predicted a Galactic AIC rate of 
$\sim 10^{-6} - 10^{-4} $ yr$^{-1}$ from population synthesis, though
note that this rate does not include mergers of CO-CO WDs.   
A Galactic rate of $\sim 10^{-4}$ AIC yr$^{-1}$ 
is the upper limit estimate derived 
from solar system abundances of neutron-rich isotopes, 
which are expected to be produced in AICs \citep{Fry99,Des06,Met09}.}, 
then this provides a potentially strong constraint
on the outcome of these mergers -- 
namely that a non-negligible fraction of SNe Ia must be formed 
through the DDS channel.  \citet{MB10} find that about half of SNe Ia
explode within 330 Myr of star formation, with a significant fraction 
occurring within 1 Gyr of star formation.  If both the DDS and
sub-Chandrasekhar mass channels contribute to SNe Ia at short delay
times, this could potentially explain both the bimodality in the
observed DTD, as well as increase the predicted number of prompt SNe Ia (see
Fig. 1).

The sub-Chandrasekhar model is the first model which demonstrates both
a sufficient number of SNe Ia events to account for all (or at least
some substantial fraction) of SNe Ia, as well as 
{\em two distinct evolutionary channels with their own characteristic
  delay times}: A prompt ($<500$ Myr) helium-star channel originating from
binaries with more massive secondaries, and a more delayed ($>500$
Myr) double WD channel originating from AM CVn-like progenitor
binaries with lower mass.  
Considering the recent observational studies  
which have found evidence for such a DTD  
\citep{MB10,Bra10,Mao10}, further theoretical investigation of the
sub-Chandrasekhar mass SN Ia channel is now strongly warranted.

%%%%%%%%%%%%%%%%%%%%%%%%%%%%%%%%%%%%%%%%%%%%
%% Sample figure:
%%
%% The option [height=...] scales the picture to the given height,
%% without it it would be printed at its nominal size
%%%%%%%%%%%%%%%%%%%%%%%%%%%%%%%%%%%%%%%%%%%%

%\begin{figure}
%  \includegraphics[height=.3\textheight]{golfer}
%  \caption{Picture to fixed height}
%\end{figure}

%%%%%%%%%%%%%%%%%%%%%%%%%%%%%%%%%%%%%%%%%%%%
%% SAMPLE TABLE
%%
%% Shows the use of \tablehead and \tablenote
%% macros
%%%%%%%%%%%%%%%%%%%%%%%%%%%%%%%%%%%%%%%%%%%%

%%%%%%%%%%%%%%%%%%%%%%%%%%%%%%%%%%%%%%%%%%%%%%%%
%% BACKMATTER
%%%%%%%%%%%%%%%%%%%%%%%%%%%%%%%%%%%%%%%%%%%%%%%%

\begin{theacknowledgments}
AJR would like to thank the organizers of the International Conference
on Binaries.
\end{theacknowledgments}

\end{document}